# TrueEBSD: correcting spatial distortions in electron backscatter diffraction maps

Vivian S. Tong**, T. Ben Britton*
Department of Materials, Imperial College London, Exhibition Road, London, United Kingdom, SW7 2AZ
*Corresponding author: b.britton@imperial.ac.uk
**Present address: National Physical Laboratory, Hampton Road, Teddington, Middlesex, TW11 0LW

## Abstract

Electron backscatter diffraction (EBSD) in the scanning electron microscope is routinely used for microstructural characterisation of polycrystalline materials. Maps of EBSD data are typically acquired at high stage tilt and slow scan speed, leading to tilt and drift distortions that obscure or distort features in the final microstructure map. In this paper, we describe TrueEBSD, an automatic postprocessing procedure for distortion correction with pixel-scale precision. Intermediate images are used to separate tilt and drift distortion components and fit each to a physically-informed distortion model. We demonstrate TrueEBSD on three case studies (titanium, zirconium and hydride containing Zr), where distortion removal has enabled characterisation of otherwise inaccessible microstructural features.

## Highlights

- TrueEBSD is a software method to correct EBSD map distortions.
- Distortion correction is a fully automatic, offline postprocessing step.
- Extra hardware is not needed to use TrueEBSD.
- TrueEBSD improves microstructural characterisation capability.

# Graphical Abstract

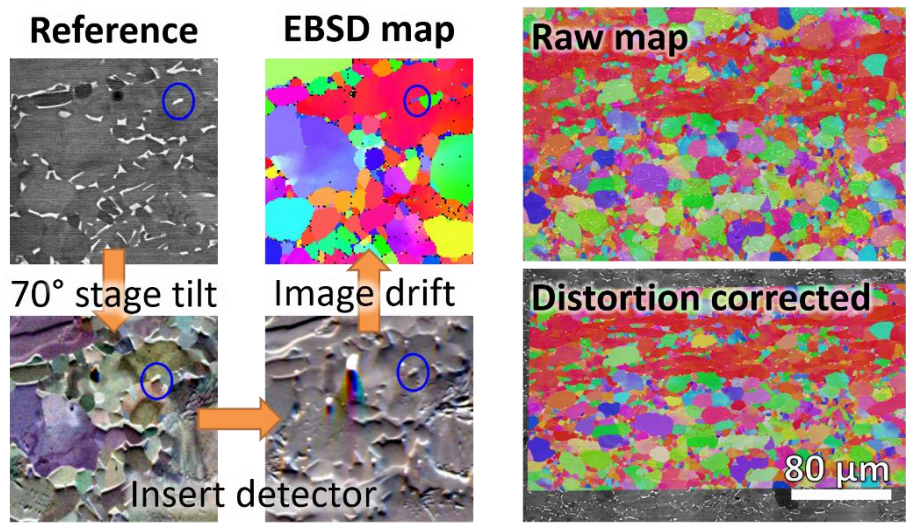

## 1. Introduction

Electron backscatter diffraction (EBSD) is a useful scanning electron microscope (SEM) based microstructural characterisation technique. Using EBSD, we can map phases and crystal orientations with high spatial and angular resolution, and the data acquisition process can be highly automated. EBSD is typically performed at high stage tilt angle as diffracted electron signal is maximised when electrons hit a sample at glancing incidence [1]. However, spatial distortions in the SEM become pronounced at high stage tilt angles, and lead to both morphology and orientation errors in EBSD maps [2,3].

Morphological and orientation distortions become problematic in correlative microscopy e.g. slip trace analysis using SEM-based digital image correlation (DIC) and EBSD, where EBSD grain orientations are required to identify the slip plane trace in as a crystal direction. In Reference [4], EBSD distortions prevented 'traditional' grain-to-grain DIC slip trace analysis, as grain boundaries could not be accurately overlaid onto DIC strain maps.

In addition to tilt distortions, EBSD mapping is typically orders of magnitude slower than electron imaging, so mechanical or thermal fluctuations in the SEM lead to a temporal drift. Gee et al. used a skew distortion to correct for temporal drift in long EBSD maps (> 16 hours) [5], assuming a constant drift velocity. Zhang et al. modelled changing drift velocity with a thin-plate spline model, a physically based interpolation scheme, but their algorithm requires manual selection of control points [6]. Charpagne et al. developed a procedure for aligning multiphase EBSD data with phase-contrast images without manual selection of control points [7]. Their distortion model used high-order polynomial functions that are not physically meaningful, but the final distortion function shows pixel-scale accuracy at phase boundaries.

Distortions are well known in EBSD analysis [8], and they are especially pronounced in 3D EBSD maps where there may be distortions associated with serial sectioning artefacts. In the present work, we focus on the distortions present only within each 2D slice. We refer the reader to other work that explores methods to improve the fidelity of 3D volume reconstructions (e.g. [9,10]).

## 2. Method

We have developed TrueEBSD which is a software postprocessing method to correct distortions in EBSD maps. No SEM hardware modifications are necessary, and the correction procedure can be run offline. TrueEBSD measures and corrects EBSD map distortion by registering microstructural features to a reference ('true') image that does not contain these distortions. This method is designed to work for polycrystalline materials, as grain boundaries (and features associated with them) are used to create a control-point free spatial correction.

In brief, TrueEBSD involves sequential measurement of a distortion model using intermediate frames. We start with a flat image, e.g. backscatter electron image, and then tilt the sample and capture multiple forwardscatter electron images as we insert the EBSD camera. The final registration is performed with respect to the raster based EBSD orientation map. Using these images and maps, an series of edge based images are generated for sub-window based cross correlation to generate the distortion field for mapping the location of each EBSD pattern (and the associated indexing) to the 'true' image frame of reference. This method is outlined in Figure 1.

For EBSD mapping, the distortion field is usually a composite of several physical phenomena and TrueEBSD uses intermediate images to enable separate measurement of each distortion type. The distortions are combined to produce the final distortion field between the reference image and EBSD map. The distortion field can be used to map the spatial coordinates of each EBSD pattern, and their associated data (e.g. phase assignment or crystal orientation), into the reference image mapping frame.

Several image registration techniques for aligning electron micrographs have been reported in the literature [6,7,11–13]. For TrueEBSD, we have optimised our method to correlate the EBSD data to the reference image grid and map the distortion field by fitting a model to a grid of local displacements. This grid is measured by image registration of rigid sub-regions. In conventional image registration using methods such as cross-correlation, matching features must have similar contrast/intensities. In the prior work, as image intensities are not preserved between SEM imaging modes images need to be appropriately filtered to achieve similar contrast modes, e.g. a Canny edge filter is used in Reference [12] to highlight crystal boundaries and topography features; alternatively, a multimodal image registration method can be used, such as maximising mutual information in a joint image histogram used in Reference [11], which is insensitive to absolute image intensities.

We present the TrueEBSD algorithm in steps as needed to map the EBSD map onto the reference data, which we take as ground truth. In most cases, we use a backscatter electron (BSE) micrograph as the ground truth, but we also show that an optical micrograph can be used. The ground truth may contain distortions, e.g. due to low magnification issues in the scanning electron micrograph, and the TrueEBSD algorithm is designed to match the mapped EBSD data to this ground truth only.

## 2.1 Algorithm inputs

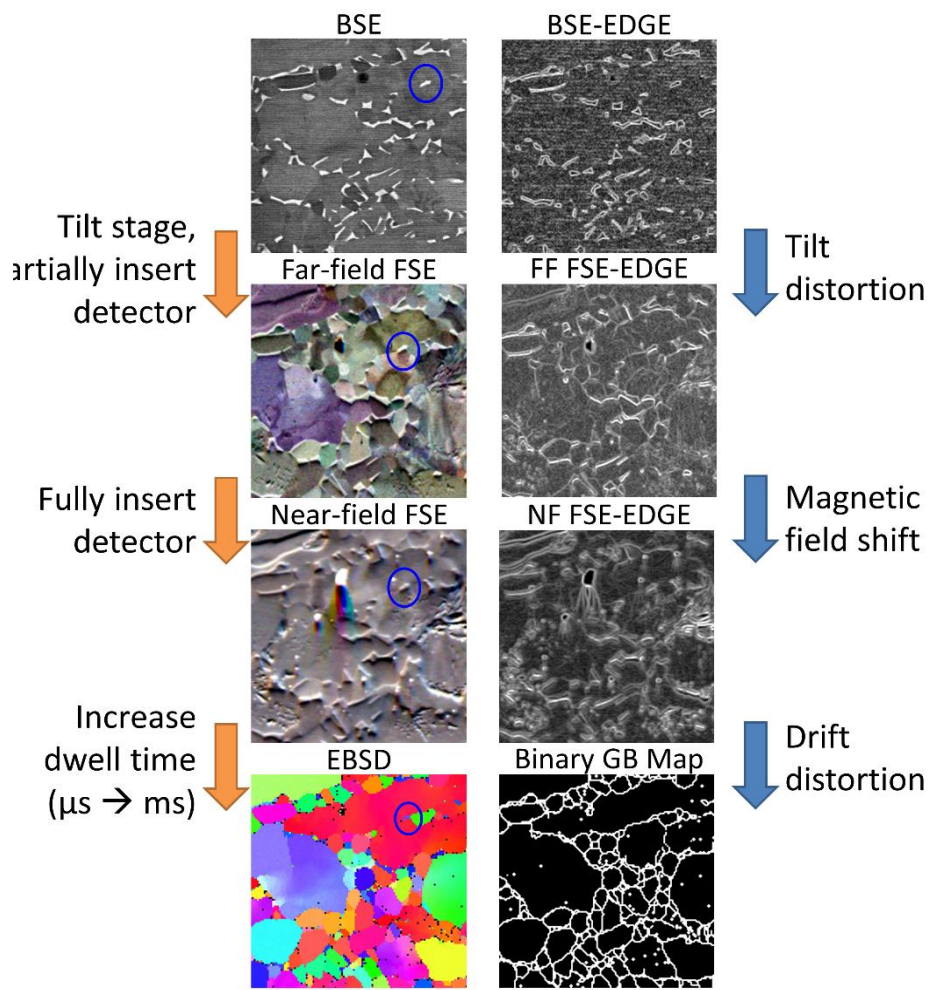

Figure 1: Imaging modes in a possible TrueEBSD workflow for a two-phase Ti alloy, and the physical origins of the distortions between them. The same microstructural feature in all the images is circled in blue. The input data is presented in the left-hand column. The edge maps are presented in the right-hand column and are used for the image correlation steps.

Within the SEM, the reference image can be an electron image taken at low sample tilt angle. Depending on the image spatial accuracy required, spatial distortions inherent to SEM imaging could be measured and corrected using a standard calibration grid, and for example a distortion correction procedure developed for SEM-based DIC strain measurements is described in Reference [14]. Note that the spatial accuracy requirement for pixel-based correlative microscopy applications is lower than for SEM-DIC strain measurement, because strain is calculated from derivatives of the displacement field.

Optical micrographs showing grain contrast, such as of etched grain structures, or polarised light images of non-cubic crystals, are also suitable. Optical micrographs captured using parallel illumination can be advantageous to avoid distortions inherent to the rolling shutter used in SEM. SEM image drift can be theoretically corrected by compositing images acquired with different fast-scan directions [15], but the distortion is usually small compared to EBSD map features.

The only requirement for the reference and intermediate images is to contain features which form a path linking the reference image to the EBSD map data. Care should be taken to maintain consistent fields of view and known scaling between nominal pixel sizes of all images. An intermediate image taken with the SEM stage tilted to EBSD configuration separates tilt from drift distortions.

Figure 1 shows a reference image, intermediate images, and EBSD map used for TrueEBSD, along with the types of distortion that link these images. The reference image is a backscattered electron image acquired at 0° stage tilt, with contrast optimised to show changes in crystal phase and orientation.

The two intermediate images are forescatter electron (FSE) images, acquired here using Bruker ARGUS forescatter imaging diodes [16] mounted to the bottom of the EBSD detector. FSE images can be acquired with the EBSD detector fully inserted ('near-field FSE') or partially retracted ('far-field FSE'). The far-field FSE image has strong electron channelling contrast [17,18], whereas the near-field FSE image has weaker electron channelling contrast and more topography contrast.

The stage is tilted, typically by 70°, between reference BSE image and far-field FSE image acquisition, so distortion between these images is because of stage movement. The SEM stage and EBSD detector do not move between near-field FSE and EBSD map acquisition, so distortion between the two is only from SEM drift, since the dwell time per image pixel increases from a few microseconds to several milliseconds [6,13]. EBSD detector insertion into the SEM chamber changes the magnetic field in the SEM chamber, deflecting the electron beam slightly between the far-field and near-field FSE images.

Each of these imaging modalities have significantly varying contrast. To provide mutual information between the modalities, we transform each image into an edge map (see right hand column of Figure 1) to reveal mutual information suitable for cross correlation between frames.

For images, the edge map is calculated per pixel by comparing the pixel with its four nearest neighbours. The value of the edge function is the sum of the square of the intensity differences for each intensity (e.g. grey or R,G,B) channel.

For the EBSD map, the edge map is calculated from a binary grain boundary map. A (white) grain boundary point is located where the neighbour orientations are greater than a threshold value (e.g. 5°).

For more information on each of these algorithms, the reader is direction to the scripts provided (see the Data Availability statement) for precise details of the edge transforms use.

## 2.2 Distortion models

### 2.2.1 EBSD detector insertion: affine transform (scale, rotation, translation, shear)

$$\begin{pmatrix} A_1 & B_1 & C_1 \\ A_2 & B_2 & C_2 \end{pmatrix} \begin{pmatrix} x_i \\ y_i \\ 1 \end{pmatrix} = \begin{pmatrix} x_i' \\ y_i' \end{pmatrix}$$

| Transformation | Matrix |
|---|---|
| X, Y scale | $\begin{pmatrix} S_x & 0 & 0 \\ 0 & S_y & 0 \end{pmatrix}$ |
| Rotation | $\begin{pmatrix} \cos\theta & \sin\theta & 0 \\ -\sin\theta & \cos\theta & 0 \end{pmatrix}$ |
| Translation | $\begin{pmatrix} 1 & 0 & t_x \\ 0 & 1 & t_y \end{pmatrix}$ |
| Shear | $\begin{pmatrix} 1 & \gamma & 0 \\ \gamma & 1 & 0 \end{pmatrix}$ |

**Equation 1: The affine transformation matrix, and decomposition into individual transformations**

To map the EBSD data back into the ground truth, we must distort the spatial grid of the EBSD data. To do this, we apply a series of transformations including the affine transformation [18], which has translation, scaling, rotation and shear degrees of freedom (Equation 1).

To create our mapping function, we need to link the different captured frames together. Between near field and far field backscatter images, our first correction is a translation as this dominates for EBSD detector insertion.

### 2.2.2 Sample tilt: projective transform

$$\begin{pmatrix} A_1 & B_1 & C_1 \\ A_2 & B_2 & C_2 \\ A_3 & B_3 & 1 \end{pmatrix} \begin{pmatrix} x_i \\ y_i \\ 1 \end{pmatrix} = \begin{pmatrix} u_i \\ v_i \\ w_i \end{pmatrix}$$

$$x_i' = \frac{u_i}{w_i} \; ; \; y_i' = \frac{v_i}{w_i}$$

**Equation 2: Projective transformation matrix.**

Next, we can consider how to best map the tilted images back onto the flat backscatter images. Here the projective transformation describes tilt distortions in SEM imaging.

There are potentially other (higher order) distortions and a detailed description of SEM tilt distortions can be found in Reference [2]. To encompass these higher order distortions, we can relax our transformation from an affine (Equation 1) to a projective (Equation 2) transform and include keystone distortions ('trapezoidal distortions' in [2]), distorting a square into a trapezium (Equation 2). We have found that keystone distortions are significant for low magnification EBSD, where large beam deflection and high sample tilt angles are required. At higher magnification, the beam deflection angle is small and affine distortions ('rhomboidal distortions' in [2]) are dominant.

Once the 70° stage tilt is accounted for, there are smaller affine tilt distortions which are magnification independent. These smaller distortions result from misalignments between the sample surface, beam scan direction, and SEM stage tilt axis.

The two extra degrees of freedom in a projective transformation can decrease the accuracy of the fitted transform. Therefore, the tilt distortion is measured in two steps: first we estimate a projective transformation, then fit any residual distortions to an affine transform.

Small (< 2-3 pixel) systematic distortions can remain after these two fitting steps. These might be accumulated from residuals in the projective and affine fits, or from other systematic instrument errors. For example, dynamic focussing improves image focussing at low magnification and high sample tilt, but dynamic focussing can also rotate the beam scan direction over the image field of view. An example of this is shown in Figure 3 of Reference [2]. As a result, parallel lines in the beam fast-scan direction of an SEM image may not be parallel on the sample surface. We remove these distortions with a (small) correction using a 2D quadratic fit (Equation 3).

### 2.2.3 Sample drift: linear spline with rigid rows

Drift distortions are mapped between the near-field FSE image coordinate frame and EBSD orientation map.

In practice, we correlate the far-field FSE image intensities mapped into the near-field FSE coordinate frame. Stronger electron channelling contrast in the far-field FSE image is better for correlation with EBSD maps, and we exploit the fact that the displacement field between the two FSE images is predictable and can be precisely determined (Section 2.2.1).

We fit the drift distortion to a linear spline in the beam slow-scan direction (rows in our data), and rigid in the beam fast-scan direction. Horizontal and vertical drift components are independent.

This method is slightly different to that previously employed by Zhang et al. [6], who used a thin-plate spline model to fit an EBSD map to a reference electron channelling contrast image at 0° tilt, as the thin plate spline is a physically realistic model for SEM drift and includes an affine transformation as a special case. Their technique used manual control point selection, allowing registration near edges of the field of view, but control point selection is time consuming, and optimal control point density depends on local distortion extent and microstructural feature size.

In contrast, TrueEBSD measures local displacements by image registration of sub-windows (i.e. a DIC type method). Large sub-windows containing more contrast features enables robust image registration, with the drawback that displacements are not accessible near the image edges. Furthermore, fitting to linear splines avoids 'flyaway' extrapolation (at the edges of the frame). We have found that a linear spline is sufficient for pixel-precise registration of the EBSD map to the reference image.

The EBSD map is registered using a grain boundary edge map generated using a threshold grain boundary misorientation (usually 5°) in MTEX.

### 2.2.4 Instrument distortions and fitting residuals: quadratic surface fit

$$x'_i = A_1 x_i^2 + B_1 y_i^2 + C_1 x_i + D_1 y_i + E_1 + F_1$$

$$y'_i = A_2 x_i^2 + B_2 y_i^2 + C_2 x_i + D_2 y_i + E_2 + F_2$$

**Equation 3: Empirical 2D quadratic model.**

After combining the displacement fields, the nominally un-distorted EBSD map can still have small, systematic displacements relative to the reference image, for which we do not have a physical explanation for. These are likely related to stray fields in the microscope room or other systematic drift issues. We empirically correct this using a 2D quadratic fit (Equation 3) to the measured shifts, weighted by the image registration normalised cross correlation peak height.

Finally, the remapped spatial map of the EBSD data can either be used 'as is' with a non uniform step size, or (as presented here) it can be remapped using a modified version of the MTEX 'gridify' function (included in the deposited code and example scripts).

## 3. Results

The versatility and robustness of the TrueEBSD method is demonstrated on three datasets: 1) eliminating seam artefacts when combining EBSD maps of a Ti-6Al-4V macrozone; 2) correlating low-magnification EBSD with polarised optical microscopy to identify hydride-forming grain boundaries in Zircaloy-4. 3) Measuring lattice rotations from constrained slip in commercially pure zirconium;

### 3.1 EBSD map stitching in Ti-6Al-4V

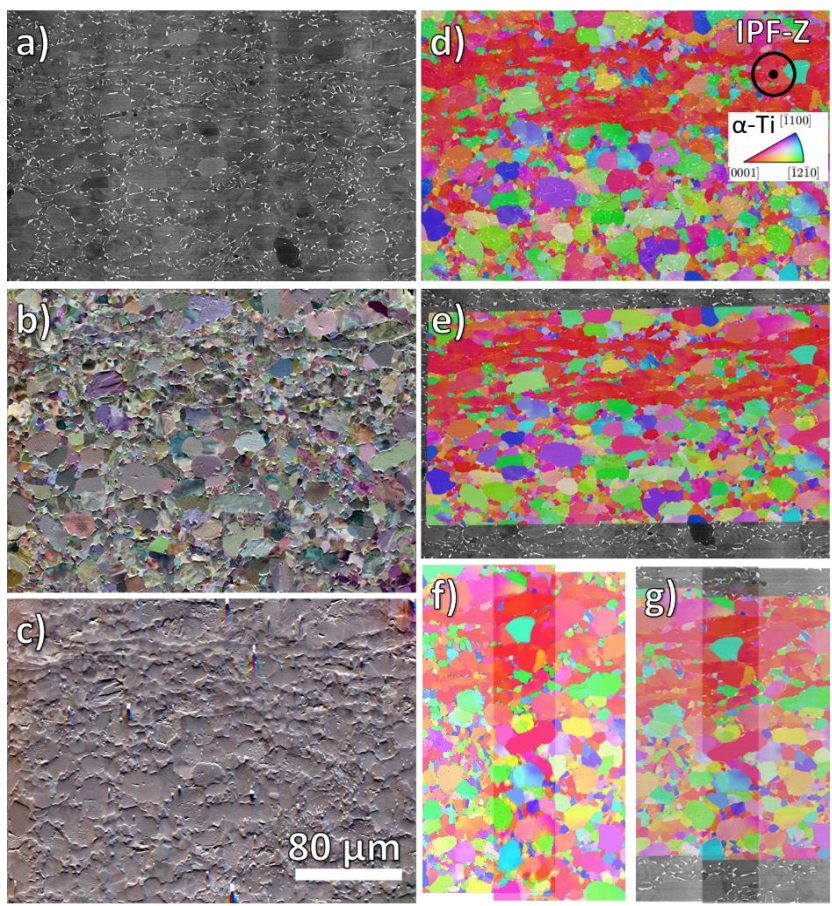

**Figure 2: Ti-64 case study: all subfigures show similar fields of view. a) Backscattered electron image of microstructure at zero tilt. b) Far field forescatter electron image showing grain contrast. c) Near field forescatter electron image in EBSD geometry. d) EBSD orientation map overlaid onto backscatter electron image showing large distortions. e) EBSD orientation map overlaid onto backscatter electron image, after distortion correction using TrueEBSD. f) Two EBSD orientation maps are taken of adjacent and overlapping fields of view. The overlapped area shows a seam artefact in the bottom half of the maps. g) Overlapped EBSD maps as in f), but with distortions corrected using TrueEBSD. The seam artefact is no longer visible.**

Page 8

EBSD mapping at high spatial resolution over large fields of view can be required to adequately characterise heterogeneous microstructures. One solution is to combine electron beam scanning based EBSD maps acquired with stage scanning, but seam artefacts from tilt and drift distortions are often visible between adjacent maps.

Distortion correction using TrueEBSD can remove the spatial component of these seam artefacts. Figure 2 shows the microstructure of a titanium alloy Ti-6Al-4V containing a macrozone, visible as a horizontal band of red grains near the top of Figure 2 (d) spanning the field of view and they often extend several hundreds of micrometres. Macrozones are detrimental to the fatigue performance of Ti alloys and can be qualitatively characterised by EBSD [19–22]. Macrozone characterisation is a potential application of stitching of EBSD maps as macrozones are much larger than the surrounding grains, but also contain grain-scale heterogeneity. Furthermore, when acquiring this dataset, TrueEBSD was also used to diagnose a mis calibrated scan card between the SEM scan card and the EBSD scan card.

For this example, the data were captured at 20 keV with 10 nA probe current. The reference image is shown in Figure 2 (a): a backscattered electron (BSE) image of Ti-6Al-4V collected at 0° stage tilt in a Quanta 650 field emission SEM [23]. Bright beta phase grains/lamella decorate equiaxed alpha grain boundaries. Figure 2 (b) and (c) show far-field and near-field FSE images respectively, collected at 70° tilt. Figure 2 (d) shows an as-acquired EBSD map overlaid on the BSE image. The distortion between the EBSD map and the reference BSE image can be seen from beta grains/lamella in the BSE image, which do not overlay onto grain boundaries in the EBSD map. Figure 2 (e) shows the same overlay after distortion correction using TrueEBSD. In addition to tilt and drift distortions, the EBSD map was artificially stretched due to miscalibration of the scan card which controls the electron beam during EBSD and FSE imaging. Figure 2 (f) shows a seam artefact between Figure 2 (d) and an adjacent EBSD map. Figure 2 (g) shows the seam artefact eliminated after TrueEBSD correction.

## 3.2 Correlating polarised light microscopy with low magnification EBSD to locate hydrides in Zr

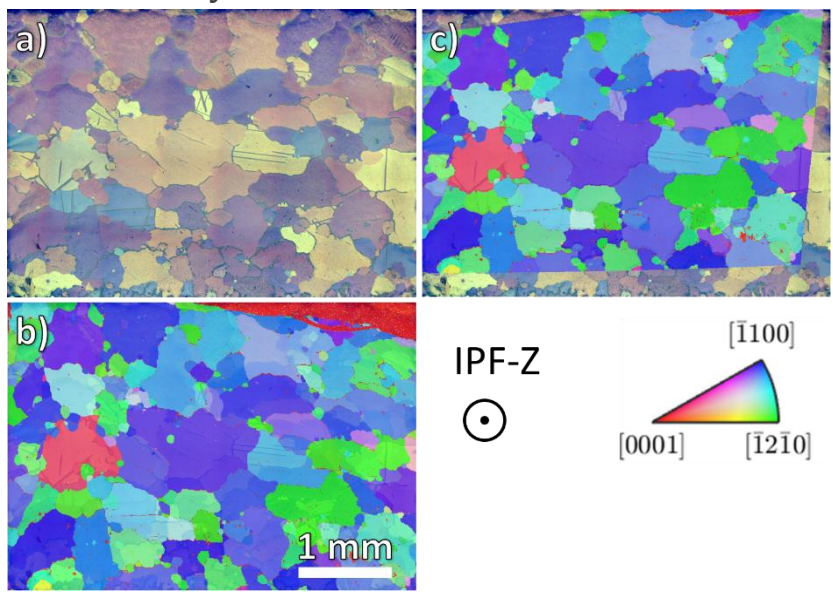

Figure 3: ZrH case study: all subfigures show similar fields of view. a) Montage of cross-polarised light micrographs, of coarse-grained Zircaloy-4 with dark hydride precipitates decorating twin and grain boundaries. b) EBSD orientation map

Page 9

overlaid onto a), showing up to ~0.1 mm boundary misalignments. c) EBSD orientation map, with distortions corrected using TrueEBSD, overlaid onto a).

The aim of this case study was to identify favourable sites for intergranular hydride precipitation in coarse-grained Zr, and results from this have been published in Reference [24]. The coarse microstructure means that the required field of view spans around 4 mm × 3mm.

The reference image in Figure 3 (a) shows a hydrided Zr microstructure stitched from 32 stitched polarised light micrographs (collected using using Olympus Stream). In this zirconium alloy sample which contains hydrides, the dark hydrides precipitate at grain boundaries, and colour contrast in the Zr grains correlates with the crystal <c> axis. The seams show colour changes from uneven illumination, but there are no spatial artefacts in this optical stitch. The electron microscopy data For this example were captured at 20 keV with 10 nA probe current. Figure 3 (b) shows the EBSD map of this region overlaid onto Figure 3 (a). In this EBSD map, Zr grains could be identified, but hydrided and unhydrided boundaries could not be reliably distinguished. The distortion is visible where grain boundaries in the optical image and EBSD map do not overlap. The purpose of this analysis was to identify preferential grain boundary locations and types for hydride precipitation. This requires use of the distortion-correct EBSD map which is overlaid onto the reference polarised light micrograph, shown in Figure 3 (c). This enabled quantitative assessment of the boundary types, as presented in Reference [24].

## 3.3 Measuring lattice rotation changes from an in-situ deformation experiment in CP-Zr

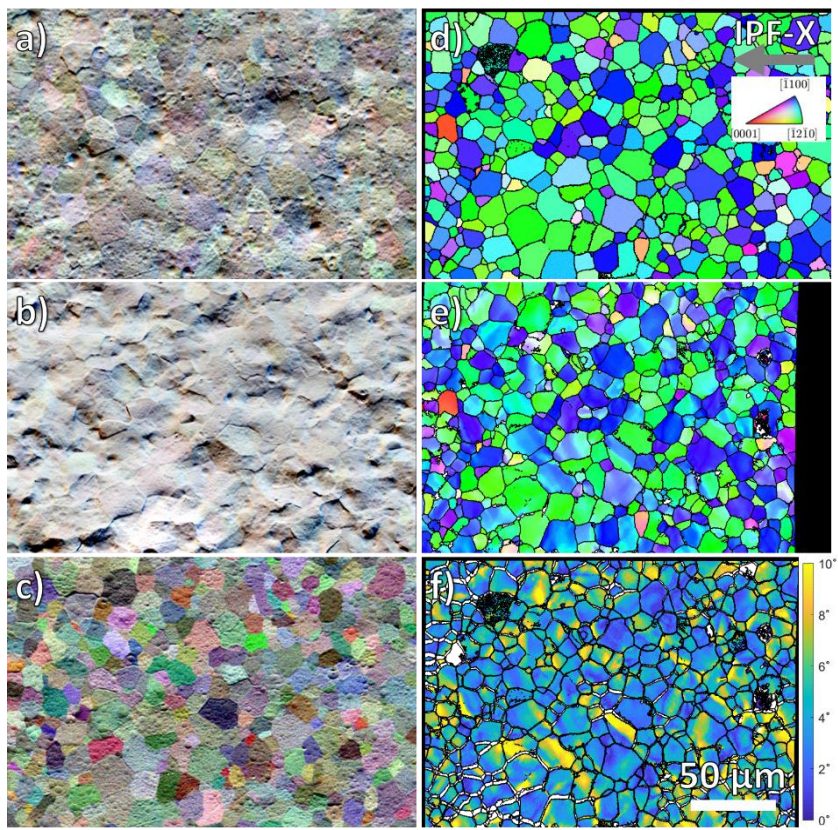

Figure 4: CP-Zr case study: all subfigures show similar fields of view. a) Undeformed sample: forescatter electron image in EBSD geometry. b) Deformed sample (12 % plastic strain): forescatter electron image in EBSD geometry immediately after unloading. c) Undeformed sample: forescatter electron image with high electron channelling contrast. d)

Undeformed sample: EBSD map registered to spatial information in c) using TrueEBSD. e) Deformed sample: EBSD map registered to spatial information in c) using TrueEBSD. The black rectangle on the right edge is from macroscopic shape change after deformation. f) Correlated EBSD map showing pairwise disorientation angles between d) and e). White regions are points with disorientation angles > 15°, i.e. regions comparing two different grains.

In-situ mechanical testing is popular to understand the evolution of plastic deformation, and EBSD measurements can help us understand signatures of plastic deformation. In particular, it is likely that changes in lattice orientation are correlated with the slip activity. Residual dislocations lead to short-range lattice curvature which can be measured using cross-correlation based and conventional EBSD [25–28], whereas long range lattice rotations are related to the degree of constrained plasticity [29–31]. Shi et al. have used quaternion image correlation combined with elastic regularisation within grains to map local plastic strain and lattice rotations in 20 % deformed Ni [32].

To aid understanding of these lattice rotations, we have used TrueEBSD to register microstructure maps before and after deformation into the same (undeformed) coordinate frame. This enables misorientation mapping with respect to the undeformed crystal orientations, and the relative contributions of macroscopic deformation and heterogeneous intergranular strain can also be assessed.

Tensile specimens suitable for a Deben MTestE2000 in-situ tensile stage [33] were machined and polished from commercially pure zirconium (CP-Zr) sheet. The final surface was polished in a Gatan PECS II argon broad ion beam polisher [34] following the method in Reference [35]. The sample was mounted onto 70° pre-tilted grips in a Quanta 650 field emission SEM [23] and loaded in tension to 1024 N, then unloaded. The gauge length measured after unloading showed 12% extension. For this example, the data were captured at 20 keV with 10 nA probe current.

EBSD mapping was performed immediately before and after deformation using Bruker Esprit 2.1 software and e-FlashHD detector [36]. Figure 4 (a) shows the near-field FSE image just before loading and Figure 4 (b) shows the same region after unloading. Surface topography in Figure 4 (b) is from grain-scale plastic rotation during deformation.

All microstructures were spatially registered to the undeformed far-field FSE image in Figure 4 (c) using TrueEBSD. Figure 4 (d) shows the drift corrected EBSD map before deformation, and Figure 4 (e) after deformation. Map colours represent inverse pole figure directions along the tensile direction (horizontal). Since the specimen was mounted on a pre-tilted sample holder, 0° tilt SEM images could not be obtained.

Figure 4 (d) shows that before deformation, the sample has equiaxed grains with negligible intragranular rotations or low angle boundaries. Figure 4 (e) shows the grains stretched in the (horizontal) tensile direction during deformation, and the black rectangle on the right shows the macroscopic strain as the map is warped to undeformed configuration. This can be measured in TrueEBSD as a scaling factor along the tensile direction. Figure 4 (f) plots disorientation angles between Figure 4 (d) and (e), which is the lattice rotation angle from slip. White 'ribbons' at some grain boundaries are where the grains in Figure 4 (d) and (e) do not match. These are likely regions that have deformed heterogeneously, which the distortion models in TrueEBSD cannot account for.

# 4. Discussion

The TrueEBSD algorithm removes spatial distortions in EBSD maps by registration to a reference image with pixel-scale precision. The measured distortion is fitted to physically relevant distortion models using intermediate images to enable distortions to be measured separately and then combined. We demonstrate good overlap of the EBSD maps to the reference data, which is explored later.

TrueEBSD is robust to relatively high levels of image noise shown in the case studies. Images with high electron channelling contrast are ideal for registration to EBSD data, but our examples show that the algorithm works even with the titanium and zirconium alloys used. These alloys do not typically have strong electron channelling contrast compared to many structural metals, and surface topography from in-situ deformation and/or sample preparation were unavoidable.

For the TrueEBSD algorithm, choice of the ground truth is important. We have found that the BSE provides a single image for the correlation, which may include low magnification distortions (depending on your microscope). The tilted images contain numerous distortions, e.g. due to misalignment of the stage, sample parallelism, and beam scanning artefacts. These are not removed reasonably in tiled map stiches, and errors may accumulate from neighbour tile effects, i.e. it is challenging to recover the ground truth from two distorted images. We note that in the ideal world, a parallel illumination with light microscopy can be beneficial, as shown in Figure 3.

In Example 1, we do not provide new quantitative insight directly in the analysis but instead highlight how the seam artefact can be properly accounted for. This enables multiple tiled regions to be selected and reduces low magnification artefacts (e.g. as shown in Example 2). In the case of the macrozone analysis, it would avoid an artificial breakup of the macrozones between tiles and enhance faithful reconstruction of a map.

## 4.1 Analysis of the residuals from TrueEBSD spatial corrections

To establish the value of the TrueEBSD algorithm presented here, we quantify the magnitude of the measured distortions in Figure 5 for all three example data sets shown here.

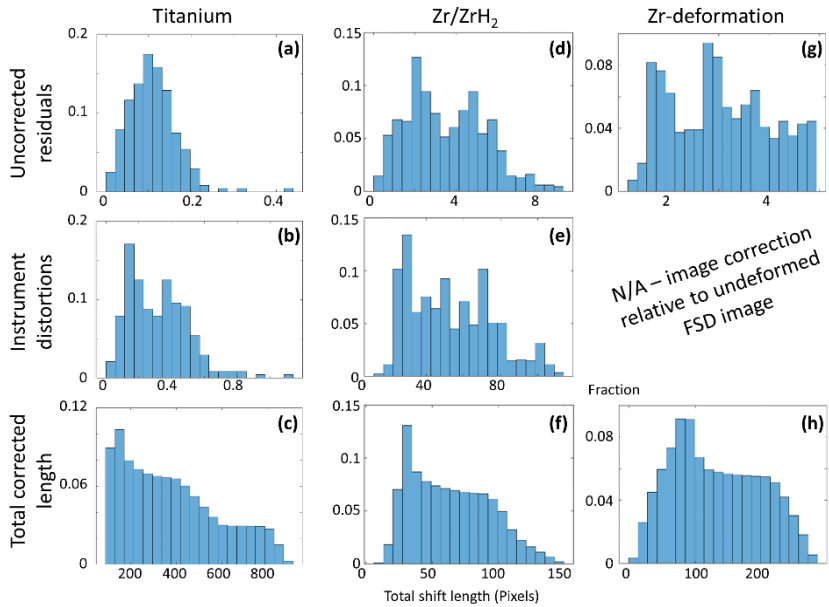

Figure 5: Digital image correlation calculated shifts for the three example datasets – Ti (a-c), Zr/ZrH2 (d-f), Zr-deformation (g,h). Highlighting the magnitude of the ultimate uncorrelated residuals, after TrueEBSD has been applied (a,d,g); the measured instrument distortions (b,e) and the total corrected length as applied to the EBSD data (c,f,h).

In the case of the titanium stitching example (Figure 5a-c), there are many features for cross correlation and model fitting. The focus of this example is the correlation of any distortions to enable seamless connection of two adjacent maps for which will enable large area mapping. The final residuals, between the EBSD edge map and the BSE edge map, tend to be less than 0.2 pixels. The maps were captured at high magnification where there are limited distortions, so the instrument corrections are also small. The degree of total correction however is large, as the two maps need to be shifted and distortion corrected for them to be connected.

In the case of the matching the large area EBSD map with the light microscopy image of the zirconium / zirconium hydride sample (Figure 5d-f), the distortions are substantial. Electron-optical distortions are known to be significant at low magnification and high tilt, yet many users prefer a 'single-shot' image as this reduces challenges associated with stitching (as noted in the titanium example). In this case, the number of grain boundaries are small which reduces the number of features that TrueEBSD uses for cross correlation and model fitting. The model was also constructed without intermediate images, so access to creation of a separate model for beam shift and low magnification tilt distortions was unavailable. This results in large residuals at the end of the model fitting (9 pixels) and yet the final map was still suitable for its ultimate purpose.

For our final case where the deformed configuration was mapped back to the initial configuration (Figure 5g-h), we are essentially removing the effect of the global plastic strain field from the spatial map of the EBSD data. Importantly, the model used to remap the global plastic strain field is more rigid than it is found in practice, as plastic strain in metals is heterogenous (especially in HCP metals like zirconium). This can be observed in the quiver plot within the supplementary data, where the strain differences between patches of grains are revealed in the residual shift maps. This results in a reasonably large residual (5 pixels). The total strain applied determines the total shift correction required (200 pixels).

## 4.2 Beyond distortion correction: other applications of TrueEBSD

The origins of EBSD spatial distortion need not be limited to microscope artefacts. For the case study in Section Figure 4, TrueEBSD is first used to remove EBSD drift distortion artefacts, and then extended to quantify microstructural distortions from sample deformation: the macroscopic strain (black rectangle in Figure 4 (e)) from scaling factors in the affine transform (Equation 1), lattice rotations from partially constrained slip (Figure 4 (f)), and heterogeneously deformed regions identified from residual displacements after fitting to homogeneous models (white regions along grain boundaries in Figure 4 (f)).

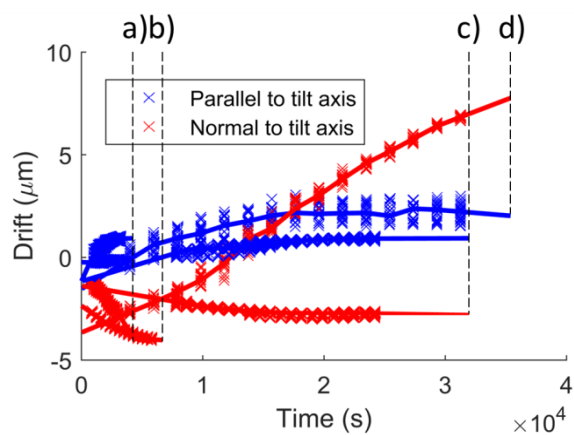

Figure 6: Drift rate for four EBSD datasets along (red) and perpendicular to (blue) the stage tilt axis. Crosses show measured displacements and lines are the linear spline fits. a) Drift rate for the undeformed CP-Zr dataset in Section 3.3. b) Drift rate for the Ti-64 dataset in Section 3.1. c) Drift rate for the deformed CP-Zr dataset in Section 3.3. Data was collected from two different instruments.

SEM drift rate can be plotted directly from drift-correction part of TrueEBSD output. This enables assessment of SEM stage or EBSD detector stability. Drift displacements for four EBSD maps of different durations are shown in Figure 6. All samples were well-grounded metallic samples to eliminate drift from sample charging. Datasets (a), (b) and (c) were acquired in a FEI Quanta field emission SEM [23], and (d) in a Zeiss Auriga field emission SEM [37]. The drift rates vary strongly between datasets and do not follow a predictable form, but the measurements are consistent with results from Reference [6] reporting velocities between 0.3 and 3 nm/s in the principle drift direction. In Figure 6 (d), there is 3 μm drift displacement in less than 10 minutes between near-field FSE imaging and starting EBSD map acquisition. Qualitative inspection of grain positions in the EBSD map and FSE image confirms that this is real, but a much higher drift rate than expected.

## 4.3 Limitations

Although TrueEBSD can correct spatial distortions in EBSD maps, high SEM tilt distorts not only spatial features but also the measured orientations, and TrueEBSD does not correct for this. The origins of tilt-related orientation distortion and a method to measure and correct this is described in Reference [2]. Orientation distortion is not usually significant for low angular resolution EBSD applications such as texture analysis, or applications which do not need absolute orientation information, such as EBSD for grain identification or grain boundary analysis.

Another limitation of TrueEBSD is that it does not consider EBSD map artefacts from spatial resolution or poor indexing. This could apply especially to nanocrystalline materials, low density and/or atomic number phases [38], beam sensitive materials, highly deformed structures, multiphase materials, and phases susceptible to orientation pseudosymmetry. The performance of TrueEBSD in the presence of significant EBSD artefacts has not been assessed. This does not rule out TrueEBSD as a useful tool, but that it may not work as a 'black box'. These artefacts could influence e.g. best sub-window size for image registration, or relative weights of measurement points when fitting a distortion model. For EBSD maps with unusable orientations, e.g. maps containing unknown phases, EBSD pattern quality could be substituted as the TrueEBSD input, as pattern quality does not depend on a correct indexing solution.

## 5. Conclusions

TrueEBSD is a software method to remove spatial distortions in an EBSD map with pixel-scale precision. It fits EBSD map features to a reference image using physically informed distortion models, using intermediate images to separate out and measure individual distortion components between the EBSD map and reference image. TrueEBSD can be used with most polycrystalline microstructures suitable for EBSD mapping and SEM imaging. The image registration process is fully automatic, which saves time and eliminates potential for operator mistakes or bias. The correction can be performed offline and the technique can be used on a standard SEM fitted with an EBSD detector and imaging detectors appropriate for a sample.

We have demonstrated this technique on three case studies: mapping plastic strains and lattice rotations from deformation in Zr, elimination of seam artefacts for large area EBSD characterisation, and identifying hydrided and unhydrided grain boundary types in coarse-grained Zr.

## 6. Acknowledgements

VT and TBB acknowledge funding from Imperial College London and HEIF through the Proof of Concept scheme and EPSRC EP/ K034332/1 (HexMat). TBB acknowledges funding of his Research Fellowship from the Royal Academy of Engineering. Electron microscopy was performed within the Harvey Flower Electron Microscopy Suite at Imperial College London and the Quanta was purchased within the Shell AIMS UTC. We would like to thank Ruth Birch for preparing the hydrided zirconium sample and acquiring the polarised light micrograph, and Chris Collins for providing the Ti-64 sample.